\documentclass{appolb}
\usepackage{epsfig}
\usepackage{pstricks}

\begin{document} 
\title{Chiral condensate and the structure of hadrons%
\thanks{talk at XXXI. Max Born Symposium {\it Three days of critical behaviour
in QCD}}%
}
\author{Jakub Jankowski$^1$\thanks{New address after October 1, 2013: 
Institute of Physics, Jagiellonian University, Reymonta 4, 30-059 Krak\'ow, 
Poland}~,
David Blaschke$^{1,2}$,
\address{
$^1$Institute of Theoretical Physics, University of Wroc{\l}aw,
Poland\\
$^2$ Bogoliubov Laboratory for Theoretical Physics,
JINR Dubna, Russia
}
}
\maketitle
\begin{abstract}
A model of hadron masses based on the quark structure 
of hadrons combined with effects of chiral dynamics is 
used to calculate the 2+1 flavour chiral condensate in the 
hadron resonance gas framework. Results are discussed 
in the context of recent lattice QCD data. Improvements
of the dynamical models of hadron structure are
suggested with the aim to estimate the strange sigma term
of the nucleon.
\end{abstract}
  
	
\section{Introduction}

Spontaneous chiral symmetry
breaking is, apart from
color confinement, the most 
important physical aspect
of strong interactions.
Virtual quarks and antiquarks 
of opposite chirality 
are attracted to each other
due to the strong interactions 
and destabilize the trivial vacuum state.
A condensate is formed which gives rise to a nonvanishing
expectation value of the bilinear fermionic
operator $\bar{\psi}\psi$. 

Once the temperature is raised 
chiral symmetry will be affected 
and eventually be restored in the deconfined
phase of QCD. 
In order to find the temperature dependence of 
the chiral condensate in the hadronic medium
we use the framework of the hadron resonance 
gas (HRG) with the partition function $Z_{\rm HRG}$.
From the generic definition follows in this case
\begin{equation}
\label{condensate}
\langle\bar{q}q\rangle = 
-\frac{\partial}{\partial m_0} T\ln Z_{\rm HRG}
=\langle\bar{q}q\rangle_0 + \sum_H \frac{\partial m_H}{\partial m_q} n_H(T)~,
\end{equation}
where the scalar densities for hadrons 
has been introduced as
\begin{equation}
n_H(T) = \frac{d_H}{2\pi^2}\int_0^\infty dkk^2
\frac{m_H}{\sqrt{m_H^2+k^2}}\frac{1}{e^{\beta \sqrt{m_H^2+k^2}}\pm 1}~,
\end{equation}
with the inverse temperature $\beta=1/T$ and the hadron degeneracy $d_H$.
The sum over hadronic states takes into account all light and
strange hadrons up to a mass $m_{\rm max}\sim2$~GeV.

In this note we explore the
effect of the quark substructure of hadrons on the melting
of the condensate for temperatures 
below the QCD critical temperature.
The results are discussed in the context
of recent first principle QCD 
simulations on the lattice.


\section{Constituent quark picture}

Some prior knowledge about 
the structure of hadrons is required to evaluate (\ref{condensate}).
The constituent quark picture (CQP) adopted recently 
\cite{Jankowski:2012ms}
assumes  that baryon and meson masses scale as 
\begin{eqnarray}
\label{eq:L1}
m_B &=& (3-N_s)M_q + N_sM_s+\kappa_B~,\\
m_M &=& (2-N_s)M_q + N_sM_s + \kappa_M~.
\label{eq:L2}
\end{eqnarray}
The quark masses in these mass formulae 
are the dynamical (constituent) ones and are denoted by $M_q$ 
for the light quarks and by $M_s$ for
the strange quark. The parameter $N_s$ measures the strangeness content
of the hadron and the quantities $\kappa_B$, $\kappa_M$ denote the state
dependent bining energies, assumed to be independent on the current quark 
masses.
For the open strange hadrons $N_s$ is simply 
the number of strange (anti)quarks in the hadron.
For hidden strange mesons 
-- such as the $\eta$ or the $h_1$ -- 
it is modified by the square modulus of the coefficient
of the $\bar{s}s$ contribution to the meson wave function.
There are two possible flavour assignments 
related to the flavor singlet and flavor octet 
structure of the hidden strange mesons.
The strangess counting parameter is
$N_s^{(0)}=2/3$ for the singlet and $N_s^{(8)}=4/3$
for the octet. 

Two further simplifying assumptions are made: 
excited states are assumed to have the same
flavour structure as their respective ground states, 
and any possible mixing between octet and singlet states 
is neglected. 
This approach obviously neglects virtual quark
loops and thus for example the strange nucleon sigma term
is strictly zero.


\section{Dynamical quark mass}

The dynamical quark mass is determined from the quark gap equation which
in the momentum representation for a fixed quark flavour reads
\cite{Roberts:1994dr}
\begin{equation}
S^{-1}_f(p) = i\gamma p + m_f + \int\frac{d^4q}{(2\pi)^4}g^2
D_{\rho\sigma}(p-q)\gamma_\rho\frac{\lambda^a}{2}S_f(q)\Gamma^a_\sigma(p;q)~.
\label{VacMassGap}
\end{equation}
We take Euclidean field theory with 
$ \left\{\gamma_\rho,\gamma_\sigma\right\} = 2\delta_{\rho\sigma}$ 
and 
$\gamma_\rho^\dagger = \gamma_\rho$, $\gamma p = \gamma_\mu p_\mu$. 
Here, $D_{\rho\sigma}(p)$ is the renormalized dressed-gluon propagator and 
$\Gamma^a_\sigma(p;q)$ is the renormalized dressed-quark-gluon vertex 
for which similar, nonperturbative Schwinger-Dyson equations exist.
In general, the analysis of these equations in the high-momentum limit, 
where due the asymptotic freedom of QCD perturbative methods apply, does 
reveal that in solving Eq.~(\ref{VacMassGap}) regularization and 
renormalization procedures are required. 
Nevertheless, additional assumptions and nonperturbative techniques are
required to obtain the nontrivial, chiral symmetry breaking and confining
solutions which characterize the low-momentum sector of QCD.
To gain further insights into the properties of this nonperturbative domain
and to obtain nontrivial, semiquantitative solutions of the quark mass gap 
equations it is legitimate to assume a model gluon propagator and
simplified vertex functions, but still obeying symmetry requirements of QCD 
such as chiral symmetry.

One of the possibilities is to adopt the local limit of a heavy-gluon
propagator 
and the rainbow-ladder ansatz for the vertex function
\begin{equation}
g^2D_{\rho\sigma}(p-q) = \frac{4\pi\alpha_{\rm IR}}{m_G^2}\delta_{\rho\sigma}~,
\hspace{1cm} 
\Gamma^a_\rho(p;q) =  \frac{\lambda^a}{2}\Gamma_\rho(p;q) 
= \frac{\lambda^a}{2}\gamma_\rho~,
\label{Model}
\end{equation} 
where $m_G = 0.8 ~\textrm{GeV}$ is a dynamically generated gluon energy scale. 
The fitted parameter $\alpha_{\rm IR}=0.9\pi$ is commensurate with 
contemporary estimates of the zero-momentum value of a running-coupling in 
QCD \cite{Chen:2012qr}.
This is a non-renormalizable four point quark vertex, similar to the NJL
model \cite{Nambu:1961tp}, providing chiral symmetry breaking but so far not 
confining. 

\begin{figure}[h!]%
\begin{center}
\includegraphics[width=.65\textwidth,height=0.5\textwidth]{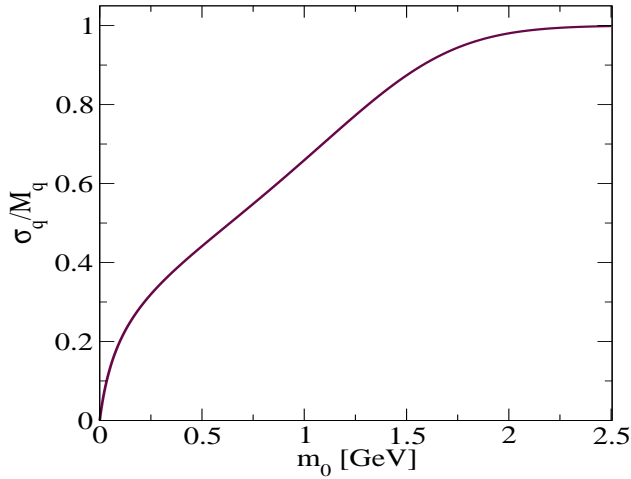}
\caption{Quark sigma term as a function of $m_0$ from Eq.~(\ref{VacMassGap}).
}%
\label{sigmaq}%
\end{center}
\end{figure}

To go beyond that, 
one of the many possibilities is to parametrize confinement by 
introding an infrared cut-off in the proper-time regularization scheme
\cite{Ebert:1996vx,Blaschke:1998ws}
and in addition to use regularization scheme preserving chiral 
Ward-Takahashi identities \cite{GutierrezGuerrero:2010md}. 
The cut in the low energy part of the 
integrals, on intuitive grounds, excludes propagation of modes
on long distances. 
Here confinement is meant to be violation of the reflection positivity 
axiom as discussed in \cite{Roberts:2000aa,Roberts:2007ji}.
Then the gap equation takes the following form
\begin{equation}
M_f = m_f + {4\alpha_{\rm IR} M_f^3}/({3\pi m_G^2})
\left[ \Gamma(-1,\tau_{uv}^2M_f^2) - \Gamma(-1,\tau_{ir}^2M_f^2)\right]~,
\label{GapVacReg}
\end{equation}
where
$\tau_{ir} = 1/\Lambda_{ir}$ and $\tau_{uv} = 1/\Lambda_{uv}$, 
with $\Lambda_{ir} = 0.24~\mbox{GeV}$ and
$\Lambda_{uv} = 0.905~\mbox{GeV}$~\cite{GutierrezGuerrero:2010md,Chen:2012qr}.
The resulting quark sigma terms are defined as
\begin{equation}
\sigma_f = m_f ({d M_f}/{dm_f})~.
\label{eq:Sigmaquark}
\end{equation}
Numerical results are shown in Fig.~\ref{sigmaq}.
It is plainly seen that dynamical effects vanish between the charm and bottom 
quark masses, i.e. around $\sim2$~GeV. 
This effect is very well known under the name of heavy quark symmetry 
\cite{Neubert:1993mb}.  
Quark sigma terms at the physical values
of current quark masses read $\sigma_q=7.65$~MeV for the up and down
quarks and $\sigma_s=121.16$~MeV for the strange quark.
The resulting nucleon sigma term is $\sigma_N=22.95$~MeV.

An easy retrospection of the hadron mass formulas shows that
the baryon octet Gell-Mann-Okubo relation
$3M_\Lambda+M_\Sigma=2(M_N+M_\Xi)$
is translated into a constraint on the state
dependent contributions:
$3\kappa_\Lambda+\kappa_\Sigma=2(\kappa_N+\kappa_\Xi)$.
Both equalities are satisfied with an accuracy of less than 5\%. 


\section{HRG and the QCD lattice data}

To compare with lattice QCD results \cite{Borsanyi:2010bp,Bazavov:2013yv} 
one considers the quantity
\begin{equation}
\Delta_{q,s}(T) = 
\left[\langle\bar{q}q\rangle-({m_q}/{m_s})\langle\bar{s}s\rangle\right]/
\left[\langle\bar{q}q\rangle_0
- ({m_q}/{m_s})\langle\bar{s}s\rangle_0\right]~.
\label{eq:}
\end{equation}
The reason to define this quantity on the lattice is purely technical: in this 
form it eliminates the quadratic singularity at nonzero values of quark mass 
$m_q/a^2$ (where $a$ is lattice spacing) and the ratio eliminates
multiplicative ambiguities in the definition of condensates. 
Physically this quantity is sensitive to chiral symmetry restoration: it is 
normalized to unity in vacuum and vanishes with vanishing 
of the condensates as temperature grows. 

The lattice results for the $\Delta_{q,s}(T)$, taken from references 
\cite{Borsanyi:2010bp,Bazavov:2013yv},
are calculated for the lattices with different temporal extent with an 
extrapolation to the continuum limit. 
The results are calculated with dynamical quark flavours for (almost) physical
values of the quark masses both for the light and the strange quarks.
The values of the quark masses are fixed by reproduction of experimental 
values of $f_K/m_\pi$ and $f_K/m_K$.


\begin{figure}[h!]%
\begin{center}
\includegraphics[width=.65\textwidth,height=0.5\textwidth]{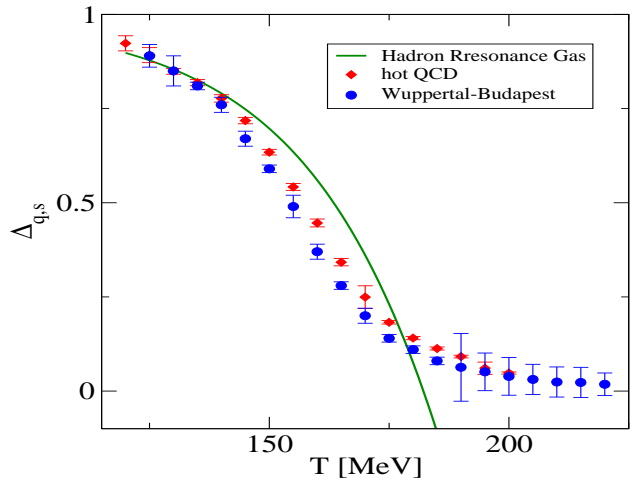}
\caption{Chiral condensate as a function of temperature calculated within 
the hadron resonance gas model (green line), compared to the lattice QCD data
of the Wuppertal-Budapest (dots) \cite{Borsanyi:2010bp} 
and hot QCD (dimonds) \cite{Bazavov:2013yv} collaborations.}
\label{Delta}%
\end{center}
\end{figure}


Fig.~\ref{Delta} shows a comparison of the lattice data 
to the HRG results with CQP mass formulas. 
It is evident that $\Delta_{q,s}(T)$ varies rapidly 
in the temperature region around $T_c\sim 155$~MeV.
There is overall agreement with HRG results up to 
temperatures $T\sim 140-150$~MeV which
is little below the pseudocritical temperature obtained 
from the lattice data \cite{Bazavov:2011nk}.


\section{Quark interactions}

To estimate effects of the interactions 
we can use the QCD version of the Breit
equation, based on taking into account
leading order relativistic corrections
to the Schr\"odinger equation \cite{Close}.
In turn the interaction Hamiltonian
has an effective potential of the form
$H^{\rm I} = H^{\rm np}+ H^{\rm SS}+ H^{\rm C}$
with the non-perturbative, spin-spin and
Coulomb parts respectively.
Instead of considering all the details of hadron
structure one can follow the logics of 
\cite{Hatsuda:1994pi} to fit the relevant
parameters with some hadron masses gaining
an overall description of sigma terms.

Once we have a potential model of hadrons virtual
quark loop effects could be estimated as a Lamb
shift of the proton mass. 
In this way the polarization loop calculated with 
pQCD can be equipped with the dynamical quark mass 
$M_s$ and the resulting screened potential  $V_{\rm sc}$
can be obtained.
In this way the mass shift is given by 
$\Delta m_p = \int~d^3r \psi^*(r)V_{\rm sc}(r)\psi(r)$
and can be related to the nucleon strange sigma term
by the Hellman-Feynman theorem.


\section{Summary}

As shown in the above note, mild assumptions
about dynamics of hadrons allow one to get 
a description of the chiral condensate at low
temperatures in a fair agreement
with the lattice data. 
Further improvments would include sigma terms 
for low lying states from chiral perturbation 
theory as those provide the dominant contributions 
to the hadron resonance gas thermodynamics.
 
The importance of the hadronic contribution
to the melting of the condensate was appreciated in a model
for the freeze-out stage of heavy ion collisions
\cite{Blaschke:2011ry}. 
In this approach freeze-out phenomena were assumed to be 
happening in the hadronic phase of QCD and were related 
to the Mott-Anderson localization of hadron wave functions 
driven by the universal chiral dynamics.
This allows for a unified explanation of the freeze-out
parameters observed in heavy ion collision experiments.

\subsection*{Acknowledgements}
We acknowledge support by the 
National Science Center (NCN) within the Maestro programme 
No. DEC-2011/02/A/ST2/00306 (D.B.), 
by a post-doctoral internship grant No. 
DEC-2013/08/S/ST2/00547 (J.J.)
and 
by the Russian Fund for Basic Research grant No. 11-02-01538-a (D.B.).

\end{document}